\documentclass{article}

\usepackage[english]{babel}
\usepackage{authblk}

\usepackage[letterpaper,top=1.5cm,bottom=2cm,left=2cm,right=2cm,marginparwidth=1.75cm]{geometry}

\usepackage{amsmath}
\usepackage{graphicx}
\usepackage{cite}
\usepackage[colorlinks=true, allcolors=blue]{hyperref}
\usepackage{amsmath}
\usepackage{amssymb}
\usepackage{booktabs} 
\usepackage{float}
\usepackage{multirow}
\usepackage[counterclockwise]{rotating}
\usepackage{caption}

\title{Identifying vulnerable nodes for hypergraph dismantling via higher-order competition dynamics}
\author[1,4]{Yajing~Hao}
\author[2,4,5,6]{Longzhao Liu\thanks{Corresponding author: longzhao@buaa.edu.cn}}
\author[2,4,5,6]{Xin Wang\thanks{Corresponding author: wangxin\_1993@buaa.edu.cn}}
\author[2,4]{Zhihao~Han}
\author[8]{Hongwei~Zheng}
\author[2,3,4,5,6,7,8]{Shaoting~Tang}

\date{}

\affil[1]{School of Mathematical Sciences, Beihang University, Beijing, 100191, China}
\affil[2]{School of Artificial Intelligence, Beihang University, Beijing, 100191, China}
\affil[3]{Hangzhou International Innovation Institute, Beihang University, Hangzhou, 311115, China}
\affil[4]{Key laboratory of Mathematics, Informatics and Behavioral Semantics, Beihang University, Beijing, 100191, China}
\affil[5]{Beijing Advanced Innovation Center for Future Blockchain and Privacy Computing, Beihang University, Beijing, 100191, China}
\affil[6]{State Key Laboratory of Complex \& Critical Software Environment, Beihang University, Beijing, 100191, China}
\affil[7]{Institute of Trustworthy Artificial Intelligence, Zhejiang Normal University, Hangzhou, 310012, China}
\affil[8]{Beijing Academy of Blockchain and Edge Computing, Beijing, 100085, China}

\begin{document}
\maketitle

\begin{abstract}
Network dismantling aims to identify a node removal sequence that can rapidly destroy network connectivity, which is an important problem for understanding the structural fragility of complex systems and designing intervention strategies. Existing studies mainly focus on pairwise networks or assume weak-deletion rules where node removal only causes hyperedges to shrink in higher-order networks. However, in many real higher-order systems, the failure of one participant may cause the entire group interaction to fail, i.e., the strong-deletion mechanism. Such a mechanism cannot be fully captured by projected networks or methods based on weak-deletion rules. To address this challenge, we propose hyper-\textbf{V}ulnerability-weighted \textbf{D}ominance rank (hyper-VDrank), a higher-order centrality method for hypergraph dismantling under strong deletion. Hyper-VDrank constructs a higher-order competition dynamics mediated by hyperedge-induced environments, where a node does not compete only with individual neighbors but responds to the collective pressure formed by other nodes in the same hyperedge. It further introduces a hyperedge vulnerability weight based on redundancy and size effects to capture the vulnerable structures, facilitating the distinction of critical nodes. Experiments show that hyper-VDrank reduces the largest connected component more rapidly, collapses the hypergraph earlier, and produces greater structural fragmentation than classical and recent methods. On 14 real-world hypergraphs, hyper-VDrank improves dismantling efficiency by 23.65\% and reduces the collapse threshold by 27.63\% on average compared with the baselines. In summary, hyper-VDrank offers an effective hypergraph dismantling tool and a new perspective on identifying vulnerable structures in higher-order complex systems.

\vspace{0.5em}
\noindent\textbf{Keywords:} complex systems; higher-order networks; hypergraph dismantling; strong deletion; vulnerable nodes; critical nodes; centrality measure
\end{abstract}

\section{Introduction}
Complex systems are ubiquitous in real-world scenarios, such as infrastructure, transportation systems, communication systems, social organizations, biological systems, and information systems~\cite{newman2003structure,boccaletti2006complex,wangprx}. These systems share a common feature: they can usually tolerate a certain level of random perturbations, but may suffer large-scale damage due to the failure of a small number of critical components~\cite{albert2000error}. Motivated by this phenomenon, network dismantling in complex networks aims to find a node removal sequence that rapidly shrinks the largest connected component and eventually collapses the network, thereby revealing the structural vulnerability of complex systems and the risks of system failures~\cite{braunstein2016network,wandelt2018comparative}. This problem gives rise to many important applications, including but not limited to immunization and intervention in epidemic or information-spreading processes, vulnerability assessment under malicious attacks, and reinforcement of critical infrastructure~\cite{wandelt2025recent}.

Node centrality in complex networks provides a major methodological foundation for network dismantling. Many studies construct node centrality measures and regard the resulting node rankings as removal sequences. Classical centrality measures include degree, betweenness, closeness\cite{albert2000error, three, eigen}, and so on~\cite{lvvital, kcore2010, pei, 2022local}. Subsequently, percolation theories such as optimal percolation and explosive percolation have further promoted the research, with representative methods such as collective influence (CI)~\cite{ci} and explosive immunization (EI)~\cite{ei}. Recently, DomiRank, a dynamics-inspired centrality method, has shown strong performance~\cite{domi}. 
It characterizes the dominance of nodes over their neighborhoods through the stationary solution of inter-node competition dynamics, and further relates such dominance to vulnerable network structures, thereby yielding a dominance-based dismantling method. 
Meanwhile, machine learning methods represented by reinforcement learning and deep learning have also given rise to approaches for network dismantling~\cite{finder,gdm,dcrs}. 

Recent studies increasingly suggest that interactions in real-world systems cannot always be fully characterized by pairwise relationships~\cite{ 23structure,23high,liu2025higher}. Processes such as group contacts, multi-agent collaboration, and scientific cooperation often involve collective interactions among multiple nodes, and thus require higher-order network representations such as hypergraphs or simplicial complexes~\cite{19high,22cp}. Higher-order interactions not only change the way network structures are represented, but also bring new challenges to network dismantling. On the one hand, higher-order networks contain more complex structural properties, such as hyperedge sizes and hyperedge overlaps~\cite{23structure}. On the other hand, the failure rule of hyperedges after node removal is diverse, making the dismantling process more complicated~\cite{strongweak,thresperco,h-percolation}. Specifically, node deletion in hypergraphs can be divided into weak deletion and strong deletion: Under weak deletion, when a node is deleted, its incident hyperedges only shrink in size; However, under strong deletion, the node and all hyperedges containing it are removed simultaneously. Such a strong-deletion mechanism naturally arises in systems whose higher-order interactions require the presence of all essential components, such as protein complexes, modular production systems, and biochemical reaction systems~\cite{reaction,hkcoreperco}. Previous studies have shown that strong deletion leads to connectivity problems different from those induced by weak deletion~\cite{strongweak}. 
Moreover, recent studies on higher-order percolation have also increasingly drawn attention to the strong-deletion mechanism, referred to as hypergraph percolation, and have shown that it can significantly change the connectivity degradation process and the critical behavior of system robustness~\cite{h-percolation,hkcoreperco}.

Recent researchers have devoted efforts to dismantling hypergraphs, such as extending classical network dismantling methods to higher-order networks~\cite{hyperci,hci} or training hypergraph-specific learning models~\cite{22reinforce-high,2026deep,hnd}. However, these methods are mostly based on weak deletion, and their effectiveness under strong deletion remains unclear. This limitation is particularly evident for learning-based methods, whose applicability highly depends on the predefined objectives used in the training process. As a result, transferring such methods to the strong-deletion setting may require a redesign of the model and training objectives. Therefore, how to identify vulnerable nodes whose removal can rapidly destroy the connectivity of hypergraphs under the strong-deletion rule remains insufficiently explored.

In this paper, we propose hyper-VDrank, a competition-based hypergraph dismantling method for identifying vulnerable nodes under the strong-deletion rule. 
Such a collective failure mechanism makes hyperedge-level structural loss central to the dismantling process. 
Therefore, we develop a higher-order competition dynamics, in which node dominance is mediated by hyperedge-induced environments instead of individual neighbors, allowing the emerging node dominance to encode higher-order structures. 
Furthermore, we equip the dynamics with a hyperedge vulnerability weight, which emphasizes low-redundancy interactions and balances the size effect of hyperedges, to embed more hyperedge-level structural information and improve performance.
Extensive experiments on synthetic and real-world hypergraphs show that hyper-VDrank consistently outperforms a wide range of classical and recent methods. On \(14\) real-world hypergraphs, hyper-VDrank achieves an average relative ANC improvement of \(23.65\%\) over the baselines and also demonstrates strong performance in terms of collapse threshold and structural fragmentation.

\section{Preliminaries and research objective}
\subsection{Hypergraph representation and node deletion rules}
\begin{figure}
    \centering
    \includegraphics[width=0.9\linewidth]{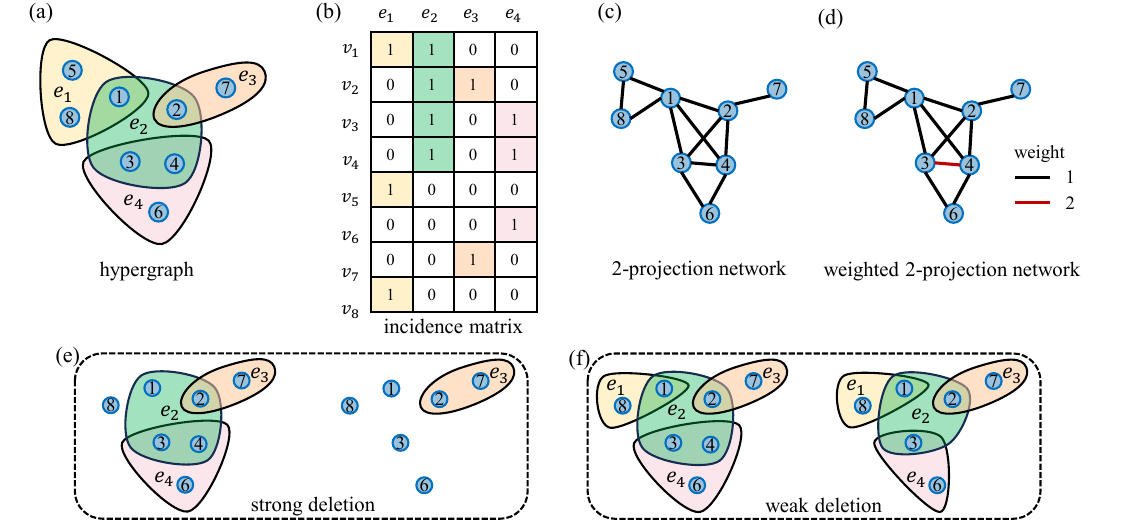}
    \caption{Illustration of hypergraph representations and deletion mechanisms. (a) A toy hypergraph with \(8\) nodes and \(4\) hyperedges. (b) The corresponding incidence matrix. 
    (c) The 2-projection network. 
    (d) The weighted 2-projection network. 
    (e) Strong deletion.
    (f) Weak deletion.
}
    \label{fig1}
\end{figure}

Given a hypergraph $H=(V,E)$ with $N$ nodes and $M$ hyperedges, where $V=\{v_1,v_2,...,v_N\}$ denotes the set of nodes and $E=\{e_1,e_2,...,e_M\}$ denotes the set of hyperedges, each hyperedge $e\in E$ is comprised of a set of nodes (i.e., $e \subset V$), as illustrated in Figure~\ref{fig1}(a). A hypergraph can be represented by an incidence matrix $B \in \mathbb{R}^{N\times M}$, where $b_{ve}=1$ if and only if node $v$ belongs to hyperedge $e$. Moreover, a hypergraph can be naturally reduced to a classical pairwise network through projection, by connecting every pair of nodes that co-occur in at least one hyperedge. For example, Figure~\ref{fig1}(c) shows the 2-projection of the hypergraph in Figure~\ref{fig1}(a). 
In addition, weighted 2-projection provides a more informative instantiation by assigning pairwise edge weights according to the number of co-occurrences of node pairs in hyperedges, as shown in Figure~\ref{fig1}(d).

The main difference between hypergraphs and ordinary networks is that their edge sizes can exceed 2, which leads to the diversity of hyperedge failure rules. The strictest rule is strong deletion, under which, once a node is removed, all hyperedges containing this node fail simultaneously. In contrast, under weak deletion, when a node is removed, its incident hyperedges remain but shrink by losing one member. Figures~\ref{fig1}(e) and \ref{fig1}(f) show the residual hypergraphs obtained by sequentially removing nodes 5 and 4 from the hypergraph in Figure~\ref{fig1}(a) under the strong and weak deletion rules, respectively. In the following, we focus on the strong-deletion rule.

\subsection{Hypergraph connectivity}
The main purpose of hypergraph dismantling is to damage the connectivity of hypergraphs. For any two nodes $u,v\in V$ in a hypergraph $H=(V,E)$, if there exists a sequence of hyperedges $e_1,e_2,\ldots,e_m\in E$ such that $u\in e_1$ and $v\in e_m$, satisfying:
\[
e_i\cap e_{i+1}\neq \emptyset,\qquad i=1,\ldots,m-1,
\]
then nodes \(u\) and \(v\) are defined as connected in \(H\). If every pair of nodes in the hypergraph is connected, then \(H\) is connected. 
A connected component of \(H\) is a maximal connected sub-hypergraph induced by a node set \(C\subseteq V\). Specifically, any two nodes in \(C\) are connected in \(H\), and maximality means that there does not exist a node \(w\in V\setminus C\) such that the sub-hypergraph induced by \(C\cup\{w\}\) is still connected. In particular, an isolated node is regarded as a component. Among all connected components of a hypergraph, the one with the largest number of nodes is called the largest connected component (LCC), which is the key indicator of the overall connectivity of a hypergraph and has received the most attention.

It is worth noting that, for weak deletion, the connectivity on the hypergraph and on its corresponding 2‑projection network are identical, because the removal of a node does not impact the connection between other neighbors.
In contrast, under strong deletion, the hypergraph dismantling problem becomes inconsistent when mapped onto the corresponding 2‑projection network, because the removal of a node is no longer local and hyperedge information is already lost in the projection. This further highlights the necessity of developing hypergraph dismantling methods specifically for strong deletion.

\subsection{Research objective}
Our objective is to develop a centrality method that determines a node removal order such that, under strong deletion, the connectivity of a hypergraph is rapidly reduced  until the hypergraph collapses, i.e., only isolated nodes remain. The dismantling performance of a centrality method is mainly evaluated using the following metrics:

\begin{itemize}
    \item LCC dismantling curve~\cite{domi}. 
    Let \(V_{LCC}(p)\) denote the node set of the largest connected component in the residual hypergraph after a fraction \(p\) of nodes is removed. The LCC dismantling curve is defined as
    \[
    LCC(p)=\frac{|V_{LCC}(p)|}{N},
    \]
    where \(N\) is the number of nodes in the original hypergraph. LCC characterizes the main connected structure of the residual hypergraph, and this curve records the entire degradation process of global connectivity during sequential node removal. A faster decrease in the curve indicates that the method can destroy the connectivity structure more effectively.
    \item Collapse threshold ($p_c$)~\cite{2026deep}. The collapse threshold is defined as the minimum removal fraction \(p\) required to completely collapse the network, namely,
    \[
    p_c=\min\left\{p: LCC(p)\le \frac{1}{N}\right\}.
    \]
    A smaller $p_c$ indicates that the method can collapse the hypergraph more rapidly.
    \item Accumulated normalized connectivity (ANC)~\cite{2026deep}. ANC is defined as
    \[
    ANC=\int_{0}^{x_c}
    \frac{\left|V_{LCC}\left(H\setminus\{v_1,\ldots,v_x\}\right)\right|}{N}\,dx ,
    \]
    where \(\left|V_{LCC}\left(H\setminus\{v_1,v_2,\ldots,v_x\}\right)\right|\) denotes the number of nodes in the largest connected component of the residual hypergraph after removing nodes \(v_1,\ldots,v_x\), and $x_c$ is the final node to remove before a hypergraph collapses.
    Equivalently, since \(p=x/N\), it can be written as
    \[
    ANC=N\int_0^{p_c} LCC(p)\,dp .
    \]
    Therefore, ANC differs from the area under the LCC dismantling curve with respect to the removal fraction only by a factor of \(N\).
    A smaller ANC indicates that the method reduces the largest connected component more effectively over the whole removal process and thus achieves higher overall dismantling efficiency.
    \item Number of connected components (CCS)~\cite{braunstein2016network,2026deep}. In addition to the above commonly used metrics for dismantling tasks, we further examine the fragmentation degree of hypergraphs during the dismantling process. Specifically, \(CCS(p)\) is defined as the number of connected components in the residual hypergraph after removing a fraction \(p\) of nodes, where each isolated node is included and counted as a component. 
    While LCC measures the scale of the largest surviving connected structure, CCS measures how many disconnected components are formed in the residual hypergraph. Hence, CCS complements LCC by quantifying the fragmentation pattern of the remaining structure. A larger \(CCS(p)\) indicates that the residual hypergraph is divided into more disconnected parts at the same removal fraction, therefore a higher degree of structural fragmentation. 
\end{itemize}

\section{Method}
The core of network dismantling is to identify nodes that play a critical role in supporting connectivity. Unlike static metrics that directly characterize topological structure, competition dynamics constructs a competitive process in which node dominance gradually emerges during the evolution toward a stationary state.
The previous study has shown that the dominance emerging from the stationary state of competition dynamics can reflect a node's influence over its neighborhood and reveal structural fragility: nodes with high dominance usually support vulnerable local structures that strongly depend on them, and their failure is therefore more likely to damage connectivity~\cite{domi}. This idea provides a basis for identifying critical nodes for dismantling from a dynamical perspective.
In ordinary networks, the core of competition dynamics is the interaction between nodes, where connected individuals compete with each other to reshape their dominance.
However, interactions are not limited to pairwise relations in hypergraphs; instead, nodes participate in collective higher-order interactions through hyperedges. 
Thus, for competition dynamics in hypergraphs, the dominance of individual nodes collectively shapes the pressure exerted by the hyperedge environment, and this environmental pressure in turn feeds back to the nodes and affects their dominance. 
Naturally, critical nodes for dismantling can be understood as those that obtain more dominance in the stationary state. In the following parts, we first extend inter-node competition to higher-order competition dynamics mediated by the collective pressure within hyperedges. 
Then, we propose a vulnerability weight to distinguish compact, low-redundancy hyperedges, thereby further facilitating the embedding of structural information into node dominance throughout the competition process. Finally, we construct the hyper-VDrank method and analyze its effectiveness for hypergraph dismantling under strong deletion.

\subsection{Higher-order competition dynamics with environmental feedback}
Let \(x_i(t)\) denote the dominance of node \(i\) at time \(t\), which reflects its competitiveness in the higher-order competition process. In a hypergraph, each hyperedge can be viewed as a higher-order environment that imposes collective pressure on its incident nodes. We assume that the dominance of a node is enhanced when the pressure from its surrounding hyperedge environments is below a competition threshold, and is suppressed when such pressure exceeds the threshold.
For each hyperedge \(e\) and incident node \(i\in e\), let \(y_{e\to i}(t)\) denote the environmental pressure exerted by \(e\) on node \(i\). This pressure is characterized by the dominance of the other nodes in the same hyperedge: if these co-participants have higher dominance, the pressure imposed on node \(i\) becomes stronger. Specifically, we formalize the higher-order competition dynamics as follows:
\begin{align}
\frac{dx_i(t)}{dt}
&=\alpha \sum_{e\ni i}w_e \left(\theta - y_{e\to i}(t)\right)-\beta x_i(t),  \label{eq1} \\
\frac{dy_{e\to i}(t)}{dt}
&=\gamma\left(\frac{1}{|e|-1}\sum_{j\in e,\, j \neq i}x_j(t)-y_{e\to i}(t)\right).  \label{eq2}
\end{align}
Here, \(\alpha\) represents the competition strength, \(w_e\) is the influence weight of hyperedge \(e\), \(\theta\) is the competition threshold, \(\beta\) is the relaxation rate of node dominance, and \(\gamma\) denotes the response rate of the environmental pressure. The first equation describes the evolution of node dominance: a node gains dominance when the environmental pressure is lower than \(\theta\), while its dominance is suppressed when the pressure is higher than \(\theta\); meanwhile, node dominance naturally diminishes at a rate of $\beta$. The second equation assumes that \(y_{e\to i}(t)\) relaxes toward the average dominance of the other nodes in \(e\), excluding node \(i\) itself, because \(y_{e\to i}(t)\) represents the pressure exerted on \(i\) by its co-participants in the same higher-order interaction. Therefore, the evolution of node dominance is driven by the indirect feedback of higher-order environments formed by hyperedges. We refer to this dynamical framework as higher-order competition dynamics with environmental feedback, or simply higher-order competition dynamics.

\subsection{Hyperedge vulnerability weight}
In the above higher-order competition dynamics, we set the hyperedge weight parameter \(w_e\) because the failure of different hyperedges may have different effects on hypergraph connectivity. Here, we design \(w_e\) from two perspectives: hyperedge redundancy and size effect. 
The goal is to incorporate structural information into the influence strength of different hyperedges in the competition process, 
thereby assisting the method to better identify critical nodes under strong deletion.

First, we define a hyperedge irreplaceability factor to distinguish low-redundancy hyperedges that play an irreplaceable role in maintaining connectivity. For highly redundant hyperedges, the relationships among their incident nodes can still be maintained through other hyperedges. In contrast, low-redundancy hyperedges are less replaceable, and their failure is more likely to destroy hypergraph connectivity, indicating higher structural vulnerability. To quantify such redundancy, we use the co-occurrence relations among nodes.
Let \(B \in \mathbb{R}^{N\times M}\) be the incidence matrix of the hypergraph, and denote the size of hyperedge \(e\) by \(d_e=|e|\). The matrix \(BB^T \in \mathbb{R}^{N\times N}\) records the number of co-occurrences between each pair of nodes. However, directly using \(BB^T\) may over-amplify the contribution of large hyperedges, since a large hyperedge naturally generates many node pairs. Therefore, we construct the following size-normalized co-occurrence matrix:
\begin{align}
    P=B\,\mathrm{diag}\left(\frac{1}{d_e-1}\right)B^T.
\end{align}
The normalization makes the total co-occurrence contribution of each hyperedge to a single incident node equal to one, since for any node \(i\in e\), we have $\sum_{j\in e,\,j\ne i}\frac{1}{d_e-1}=1$.
Thus, \(P\) measures pairwise co-occurrence support on a comparable scale across hyperedges of different sizes.
Based on \(P\), we measure the redundancy of hyperedge $e$ (denoted as $\rho_e$) through the average amount of additional co-occurrence support among node pairs inside hyperedge \(e\), which is given as:
\begin{align}
    \rho_e=
    \frac{
    \sum_{\substack{i,j\in e\\ i<j}}
    \left(P_{ij}-\frac{1}{d_e-1}\right)}
    {\binom{d_e}{2}}.
\end{align}
Here, \(P_{ij}-1/(d_e-1)\) represents the alternative co-occurrence support between nodes \(i\) and \(j\) provided by other hyperedges, excluding the contribution introduced by hyperedge \(e\) itself, namely \(1/(d_e-1)\). If node pairs within a hyperedge frequently co-occur in other hyperedges, this indicates that the hyperedge has higher redundancy and thus lower irreplaceability. Accordingly, we define the hyperedge irreplaceability factor as
\begin{align}
    q_e=\frac{1}{1+\rho_e}.
\end{align}
A smaller redundancy \(\rho_e\) leads to a larger irreplaceability factor \(q_e\), meaning that low-redundancy hyperedges are assigned higher irreplaceability.

Then, we further consider the size effect of hyperedges under the strong-deletion mechanism. Under strong deletion, a hyperedge fails once any of its incident nodes is removed. Therefore, a large hyperedge has more incident nodes through which it can be removed, whereas a small hyperedge with low redundancy has fewer such opportunities and may require more precise targeting. Motivated by this observation, we introduce a compactness preference factor \(1/(d_e-1)\). Finally, the vulnerability weight of hyperedge \(e\) is defined as
\begin{align}
    w_e=\frac{q_e}{d_e-1}.
    \label{eq3}
\end{align}
The compactness preference does not imply that large hyperedges are unimportant; rather, it balances hyperedge irreplaceability with the size effect. Overall, this weight jointly accounts for hyperedge irreplaceability and hyperedge size, allowing low-redundancy and structurally supporting hyperedges to exert stronger influence in the competition dynamics and assisting the proposed method in identifying critical nodes under strong deletion.

\subsection{Hyper-VDrank centrality}
Hyper-VDrank centrality is defined as the stationary solution of the higher-order competition dynamics with vulnerability weight.
When the dynamic system reaches the stationary state, we have
\[
\frac{dx_i(t)}{dt}=0,\qquad 
\frac{dy_{e\to i}(t)}{dt}=0.
\]
Then, by substituting the stationary condition into Eqs.~\eqref{eq1}--\eqref{eq2} and denoting $x_i(t)=x_i$, we obtain
\begin{align}
    \alpha \sum_{e\ni i} w_e 
    \left(
    \theta-\frac{1}{|e|-1}
    \sum_{\substack{j\in e,\  j\neq i}} x_j
    \right)
    -\beta x_i=0 .
    \label{eq4}
\end{align}
We now write the above equation in matrix form. First, Eq.~\eqref{eq4} can be rewritten as:
\begin{align}
    \alpha\theta\sum_{e\ni i}w_e
    -
    \alpha
    \sum_{e\ni i}
    \sum_{j\in e,\,j\ne i}
    \frac{w_e}{|e|-1}x_j
    -\beta x_i=0.
    \label{eqex}
\end{align}
Then, define
$
    M_{ij}
    =
    \sum_{e\supseteq\{i,j\}}
    \frac{w_e}{|e|-1}
$ when $i\ne j$ and $M_{ii}=0$.
We have 
$
    \sum_{e \ni i}\sum_{j \in e,j\neq i}\frac{w_e}{d_e-1}x_j=\sum_j^{N}M_{ij}x_j
$.
Then define $k_i=\sum_{e\ni i}w_e$, and the above Eq.~\eqref{eqex} becomes
\begin{align}
    \alpha\left(\theta k_i-\sum_{j=1}^N M_{ij}x_j\right)-\beta x_i=0.
\end{align}
By defining $\boldsymbol{x}=(x_1,\ldots,x_N)^T$, $\boldsymbol{k}=(k_1,\ldots,k_N)^T$, and $M$ as the matrix with entries $M_{ij}$, we obtain the following matrix form:
\begin{align}
    \alpha(\theta \boldsymbol{k} -M\boldsymbol{x})-\beta \boldsymbol{x}=0 .
\end{align}
Denote $\sigma=\frac{\alpha}{\beta}> 0$.
The closed-form stationary solution is given by
\begin{align}
    \boldsymbol{x}^*=\sigma\theta(I+\sigma M)^{-1}\boldsymbol{k}.
    \label{eq5}
\end{align}
Here, \(\theta\) only rescales \(\boldsymbol{x}^*\) and does not affect the node ranking; therefore, we set \(\theta=1\). The parameter \(\gamma\) only affects the dynamic convergence rate of the environmental pressure and does not enter the stationary solution. Thus, the node ranking produced by hyper-VDrank is independent of \(\gamma\). Therefore, once \(\sigma\) is given, the hyper-VDrank centrality for a hypergraph will be fully determined. 
To guarantee the existence of the stationary solution, \(I+\sigma M\) needs to be nonsingular, and this leads to a constraint on $\sigma$. Since \(M\) is symmetric, let \(\lambda_i\) \((i=1,\ldots,N)\) denote the eigenvalues of \(M\). The nonsingularity condition is $1+\sigma\lambda_i\neq 0, \forall i$.
Denote by \(\lambda_{\min}\) the minimum eigenvalue of \(M\). Since \(M\) is a nonzero real symmetric matrix with a zero diagonal, we have \(\mathrm{tr}(M)=0\), and thus \(\lambda_{\min}<0\). As $\sigma >0$, a continuous feasible interval is $\sigma\in\left(0,-\frac{1}{\lambda_{\min}}\right)$. In the implementation, following previous work~\cite{domi}, we search for the optimal parameter \(\sigma^*\) that minimizes ANC within the above feasible interval for each hypergraph. Specifically, if \(r\) evenly spaced candidate values are selected to form \(S_r\), then the optimal parameter is:
$\sigma^*=\arg\min_{\sigma\in S_r}\mathrm{ANC}(\sigma)$.
In practice, we use \(100\) evenly spaced candidate values from $\left[0,-\frac{1}{\lambda_{\min}}\right)$, where \(\sigma=0\) corresponds to a degenerate case with identical node scores and is included only as a reference point.

\begin{figure}
    \centering
    \includegraphics[width=0.9\linewidth]{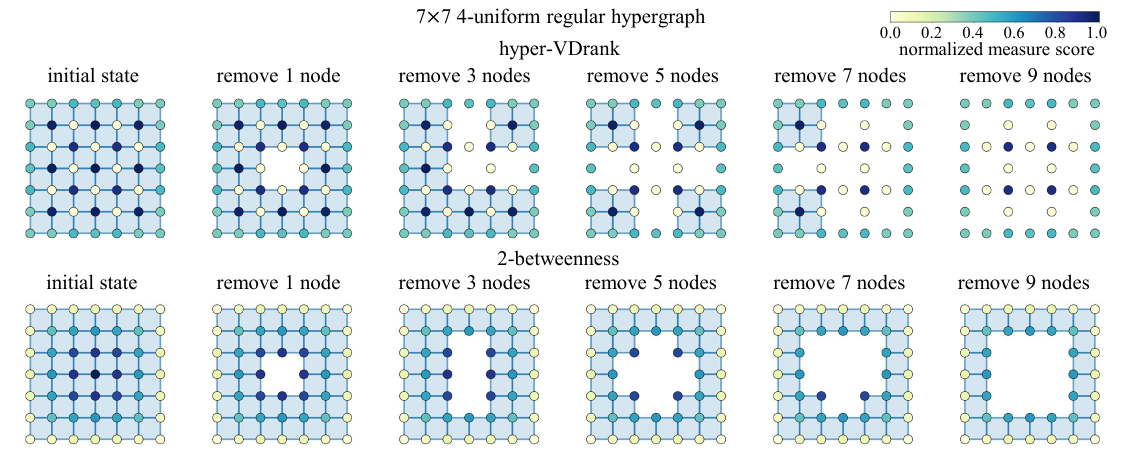}
    \caption{Dismantling process of the \(7\times 7\) 4-uniform regular hypergraph by hyper-VDrank (the upper row) and 2-betweenness (the lower row) under strong deletion. Node colors indicate the normalized measure scores assigned by the corresponding method, with darker colors representing higher scores and therefore earlier removal priority.
    }
    \label{fig3}
\end{figure}

Finally, we show that the dynamics-based hyper-VDrank enables a globally coordinated dismantling sequence to emerge spontaneously. Taking the \(7\times 7\) 4-uniform regular hypergraph in Figure~\ref{fig3} as an example, as the number of removed nodes increases, the network is rapidly divided into multiple local regions and eventually collapses. In contrast, the competitive baseline 2-betweenness, which relies on shortest-path structures, is difficult to generate such a globally joint dismantling effect.

\subsection{Comparison with projection-based competition methods}
\begin{figure}[!htb]
    \centering
    \includegraphics[width=0.9\linewidth]{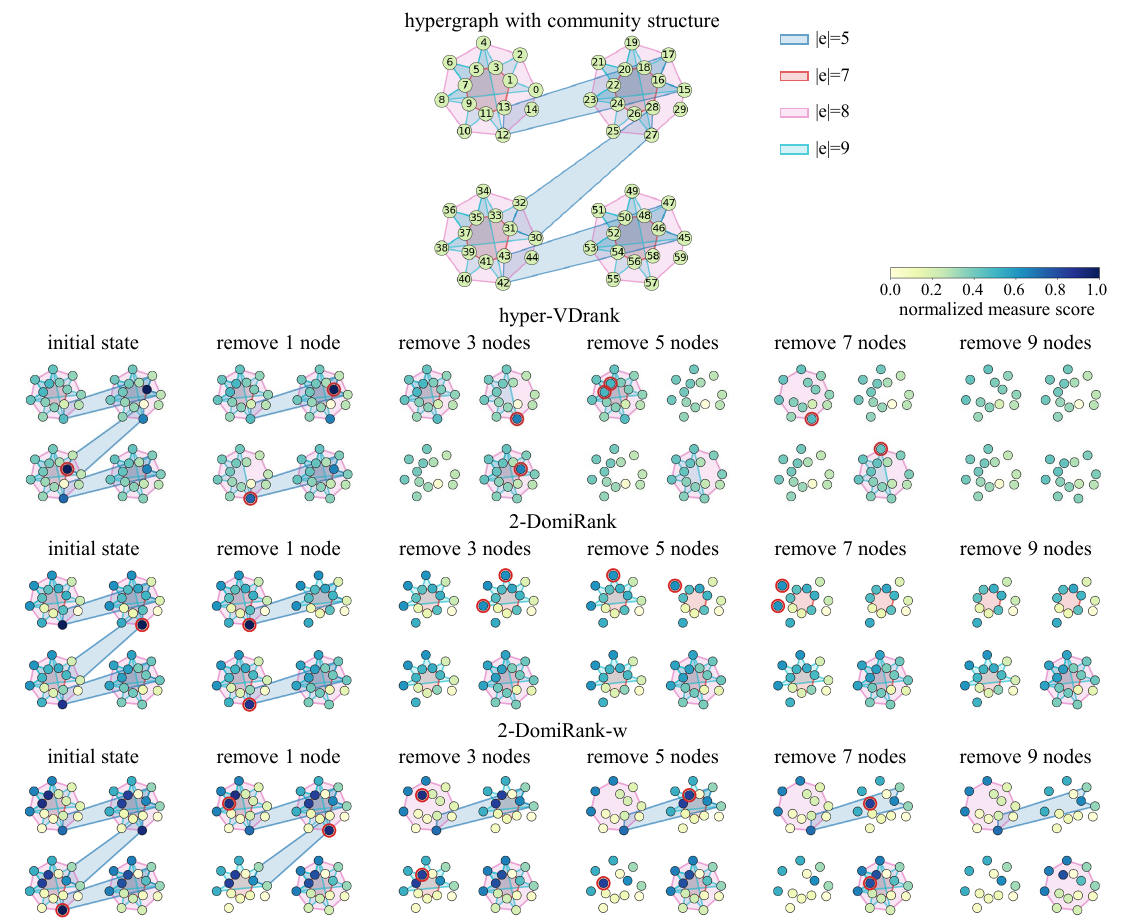}
    \caption{
    Illustration of a hypergraph with community structure and its dismantling processes under different methods. The top panel shows a synthetic hypergraph with community structure, where nodes are divided into four densely connected communities and inter-community connections are formed by hyperedges spanning different communities. Taking the upper-left community as an example, it contains a size-\(7\) hyperedge \(\{1,3,5,7,9,11,13\}\), a size-\(8\) hyperedge \(\{0,2,4,6,8,10,12,14\}\), and two size-\(9\) hyperedges \(\{0,1,2,3,4,5,6,7,8\}\) and \(\{4,5,6,7,8,9,10,11,12\}\). The connection rules in the other communities are similar. Hyperedges are colored according to their sizes. The following rows show the residual hypergraphs generated by hyper-VDrank, 2-DomiRank, and 2-DomiRank-w after removing different numbers of nodes under the strong-deletion rule. Node colors indicate the normalized measure scores assigned by the corresponding method, and red circles mark the nodes that will be removed at the next step. Compared with projection-based methods, hyper-VDrank better preserves hyperedge-level structural information and generates a more coordinated dismantling sequence under strong deletion.
    }
    \label{fig4}
\end{figure}
We further analyze hyper-VDrank by comparing it with projection-based competition methods.
From the perspective of the stationary solution, hyper-VDrank equivalently induces a weighted pairwise interaction matrix \(M\).
However, the entries of $M$ are not obtained by simply projecting hyperedges into ordinary edges. Instead, they are generated by higher-order competition dynamics mediated by hyperedge-induced environmental feedback, together with the redundancy-based hyperedge vulnerability weight. Therefore, the resulting matrix $M$ preserves information about hyperedge size, overlap, and irreplaceability, which cannot be fully retained by ordinary projection-based representations.

For comparison, we project the hypergraph into a pairwise network and then directly apply the pairwise competition-based centrality DomiRank~\cite{domi}. We refer to the unweighted and weighted projection-based variants as 2-DomiRank and 2-DomiRank-w, respectively. 
Then, we consider a hypergraph with community structure, as shown in Figure~\ref{fig4}, to illustrate the above difference. This hypergraph contains 60 nodes and 19 hyperedges, and is divided into four densely connected communities. Each community contains 15 nodes and includes one hyperedge of size 7, one hyperedge of size 8, and two hyperedges of size 9. The communities are connected by hyperedges of size 5. Figure~\ref{fig4} shows the dismantling processes of hyper-VDrank, 2-DomiRank, and 2-DomiRank-w on this hypergraph. It can be seen that hyper-VDrank tends to destroy the connections between communities and rarely removes nodes that have already become isolated, leading to stronger fragmentation and faster collapse of the hypergraph. In other words, hyper-VDrank identifies critical nodes associated with structurally vulnerable parts of the hypergraph. In contrast, the projection-based methods, 2-DomiRank and 2-DomiRank-w, lose higher-order information and may remove nodes that have already become isolated under strong deletion, making it difficult to form a joint dismantling process.

\section{Experiments and results}
To further evaluate the performance of hyper-VDrank in hypergraph dismantling, we compare it with several types of baseline methods, including classical centrality measures computed on the 2-projection network, namely 2-degree, 2-closeness, and 2-betweenness, as well as recently proposed higher-order network centrality methods, including degree~\cite{23structure}, hyper-coreness-\(R_w\)~\cite{hypercore}, eigenvector-linear~\cite{hypereigen}, HCI\(_1\), HCI\(_2\)~\cite{hci}, HDF, and EHDF~\cite{hdf}.
Degree counts the number of hyperedges incident to a node (also called hyperdegree).
Hyper-coreness-\(R_w\) is a node importance measure based on hyper-core decomposition, and it evaluates node importance according to the core-periphery position of nodes in hypergraphs. Eigenvector-linear is an eigenvector centrality measure constructed from node-hyperedge incidence relations, reflecting the spectral influence of nodes in the overall higher-order incidence structure. The HCI methods are collective-influence maximization methods based on optimal percolation theory for hypergraphs. HCI\(_1\) and HCI\(_2\) correspond to different orders of higher-order neighborhood information. To match our strong-deletion rule, we set the threshold parameter in HCI\(_1\) and HCI\(_2\) to $\frac{1}{|e|_{\max}}$.
HDF and EHDF are higher-order distance-based fuzzy centrality methods customized for hypergraphs. They measure node influence by considering the distribution of neighboring nodes within different higher-order distances.
Each method derives a node ranking sequence for node-by-node removal, which is then evaluated under the strong-deletion rule.

\subsection{Dismantling synthetic hypergraphs}
\begin{figure}
    \centering
    \includegraphics[width=0.9\linewidth]{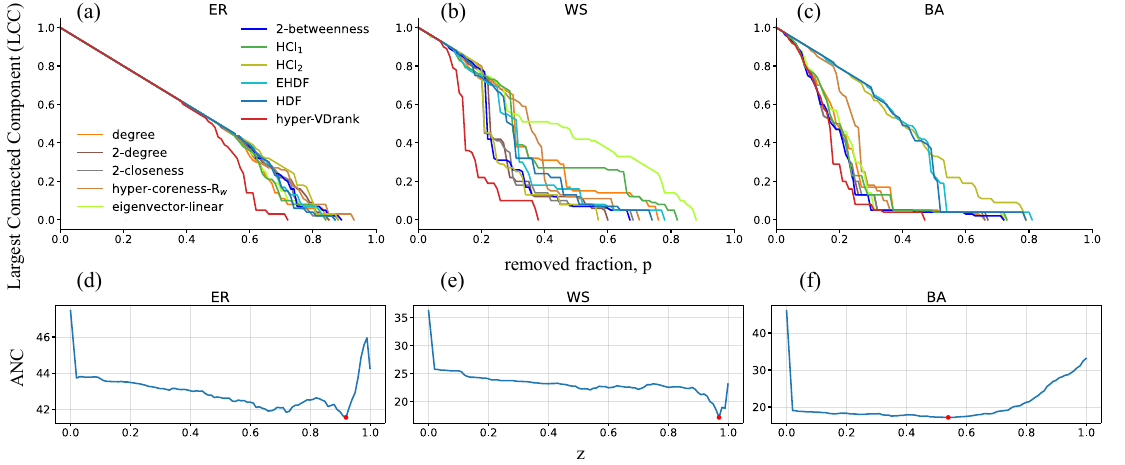}
    \caption{LCC dismantling curves and parameter search process on synthetic hypergraphs. 
    (a)--(c) LCC dismantling curves of different methods on ER, WS, and BA hypergraphs, respectively. 
    (d)--(f) ANC values of hyper-VDrank under different normalized parameters \(z=-\sigma\lambda_{\min}\) on the corresponding hypergraphs, where the red marker indicates the selected optimal parameter. 
    }
    \label{fig5}
\end{figure}

We first evaluate the dismantling performance of hyper-VDrank on three synthetic hypergraphs generated by representative hypergraph models, including ER, WS, and BA hypergraphs~\cite{er,ws,xgi,hypergraphx,han2024probabilistic}. The basic properties of these hypergraphs are given in Table~\ref{datasets}, where $m_{\text{max}}$ and $\langle m \rangle$ are the maximum and average hyperedge sizes, respectively; $k_{\text{max}}$ and $\langle k \rangle$ are the maximum and average node degrees, respectively. Figure~\ref{fig5}(a)--(c) shows the LCC dismantling curves of different methods on the ER, WS, and BA hypergraphs. For each hypergraph, nodes are removed one by one according to the ranking generated by each method. The curves are truncated when the hypergraph collapses, i.e., when only isolated nodes remain. Overall, hyper-VDrank achieves the best performance on all three hypergraphs, suggesting that it can effectively identify nodes whose removal strongly affects hypergraph connectivity under the strong-deletion rule.
On both the ER and WS hypergraphs, the LCC dismantling curve of hyper-VDrank decreases significantly faster than those of the baseline methods. Particularly on the ER hypergraph, where the hypergraph structure is relatively homogeneous and local differences among nodes are small, the baseline methods show similar performance, whereas hyper-VDrank still produces a more effective removal sequence.
In the BA hypergraph, due to the strong structural heterogeneity, the dismantling task becomes relatively easier, and several baseline methods, such as 2-betweenness, 2-closeness, and 2-degree, can also dismantle the hypergraph effectively. Nevertheless, hyper-VDrank shows outstanding performance.

\begin{table}
\footnotesize
\caption{Structural statistics of synthetic and real-world hypergraphs.}
\centering
\tabcolsep 22pt
\begin{tabular*}{\textwidth}{ccccccc}
\toprule
Dataset & $N$ & $M$ & $m_{\max}$ & $\langle m\rangle$ & $k_{\max}$ & $\langle k\rangle$ \\
\hline
ER    &100    &1516    &6     &4.50      &91     &15.16\\
WS     &100    &300    &5    &4.99    &18     &3.00\\
BA     &100    &880   &6     & 3.82    &254    &8.80 \\
Thiers13           & 327    & 4795   & 7   & 3.09 & 131   & 14.66 \\
SFHH               & 403    & 6398   & 10  & 2.72 & 222   & 15.88 \\
LyonSchool         & 242    & 10848  & 10  & 4.05 & 438   & 44.83 \\
LH10               & 76     & 1102   & 7   & 3.45 & 205   & 14.50 \\
InVS15             & 217    & 3279   & 10  & 2.77 & 141   & 15.11 \\
Elem1              & 339    & 20940  & 16  & 4.73 & 1052  & 61.77 \\
email-EU           & 979    & 24399  & 25  & 3.49 & 910   & 24.92 \\
email-Enron        & 143    & 1459   & 37  & 3.13 & 117   & 10.20 \\
senate-committees  & 282    & 301    & 31  & 17.57 & 61   & 1.07 \\
senate-bills       & 294    & 21721  & 99  & 9.90 & 3222  & 73.88 \\
music-review       & 1104   & 685    & 83  & 15.30 & 127  & 0.62 \\
geometry-questions & 580    & 888    & 230 & 13.00 & 227  & 1.53 \\
algebra-questions  & 420    & 979    & 107 & 7.57 & 328   & 2.33 \\
house-committees   & 1290   & 335    & 81  & 35.25 & 44   & 0.26 \\
\bottomrule
\end{tabular*}
\label{datasets}
\end{table}

Figures~\ref{fig5}(d)--(f) show the parameter search process of hyper-VDrank on different hypergraphs. We represent \(\sigma\) using a normalized parameter \(z\) over the continuous feasible interval, namely $\sigma=-\frac{z}{\lambda_{\min}},\ z\in[0,1)$. When \(z=0\), all nodes have hyper-VDrank scores equal to zero, so the removal order is determined randomly. The vertical axis in the figures represents ANC, and the red dots correspond to the optimal parameters. It can be seen that the optimal parameter varies across hypergraphs, suggesting that the optimal competition strength depends on the underlying hypergraph structure.

\subsection{Dismantling real-world hypergraphs}
We further evaluate the dismantling performance of different methods on \(14\) commonly used real-world hypergraphs. These hypergraphs cover a wide range of scenarios, including contact hypergraphs, email hypergraphs, committee hypergraphs, question-answering hypergraphs, review hypergraphs, and co-participation hypergraphs. Specifically, the 14 hypergraphs include Thiers13~\cite{Thiers13}, SFHH~\cite{InVS15_2_SFHH}, LyonSchool~\cite{Lyon,datas}, LH10~\cite{LH10}, InVS15~\cite{InVS15_1,InVS15_2_SFHH}, Elem1~\cite{Elem1_Mid1}, email-EU, email-Enron~\cite{datas,email-eu,email-enron}, house-committees, senate-committees, senate-bills\cite{house,bills1,datas,housesenata}, music-review~\cite{music}, geometry-questions, and algebra-questions~\cite{questions}. We first follow previous work~\cite{hypercore} to preprocess these hypergraph datasets, and then remove duplicates and take the LCC.
These hypergraphs cover many different statistics (Table~\ref{datasets}) for a broad test, including the number of nodes, the number of hyperedges, hyperedge sizes, and node degrees, and therefore offer a comprehensive evaluation in real-world hypergraphs.
For hyper-VDrank, the detailed search process of the optimal \(\sigma^*\) on each real-world hypergraph is provided in Appendix~A.

\begin{figure}
    \centering
    \includegraphics[width=\linewidth]{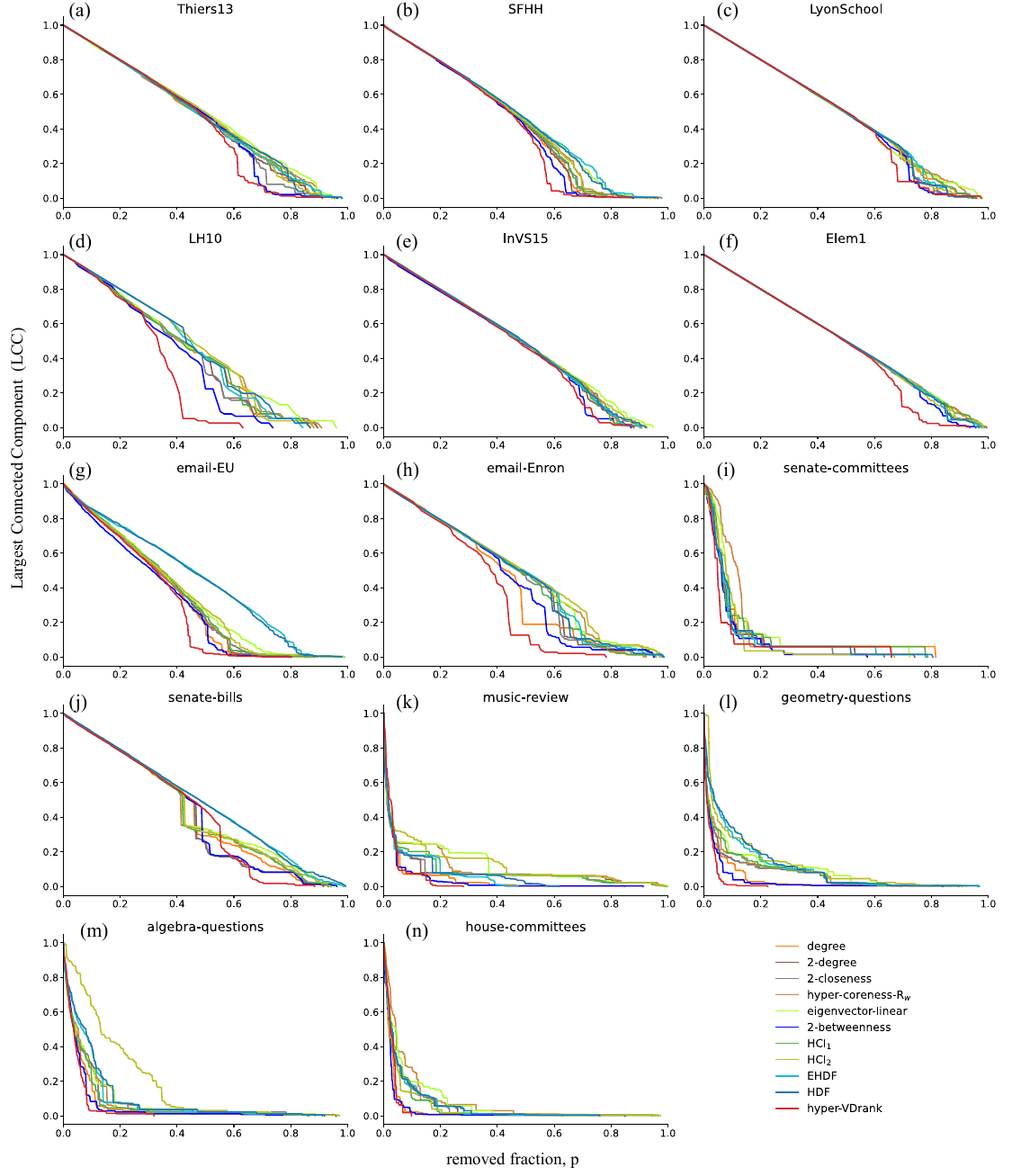}
    \caption{LCC dismantling curves on real-world hypergraphs. Hyper-VDrank reduces the LCC and achieves complete collapse more rapidly on most real-world hypergraphs, showing its effectiveness in identifying vulnerable nodes under strong deletion.}
    \label{fig6}
\end{figure}

We conduct node-by-node strong deletion dismantling on \(14\) real-world hypergraphs and Figure~\ref{fig6} shows the LCC dismantling curves of different methods.  
Overall, hyper-VDrank reduces the size of the largest connected component more rapidly on most hypergraphs, indicating its effectiveness in identifying critical nodes associated with vulnerable hypergraph structures. 
Quantitatively, hyper-VDrank achieves average relative improvements of \(23.65\%\) and \(27.63\%\) in ANC and collapse threshold \(p_c\) over the baseline methods, respectively. The detailed values of ANC and \(p_c\) are summarized in Appendix~B.

\begin{figure}
    \centering
    \includegraphics[width=0.98\linewidth]{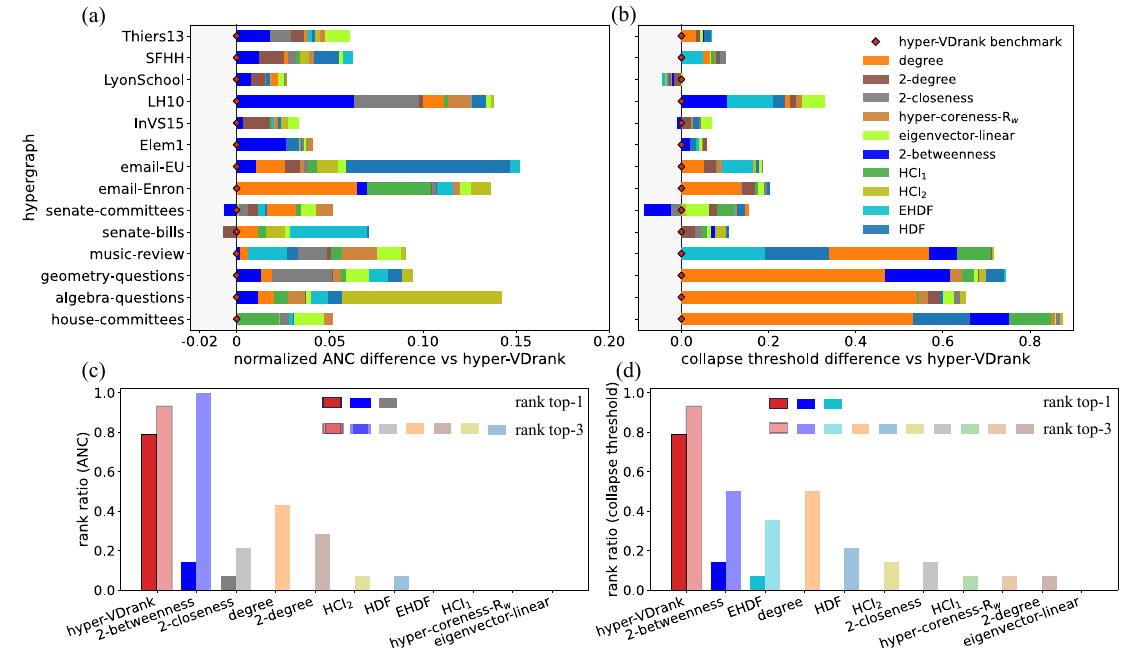}
    \caption{Quantitative comparison of performance of various methods. (a) Normalized ANC difference of each baseline method relative to hyper-VDrank. (b) Collapse threshold difference of each baseline method relative to hyper-VDrank. (c) Rank-1 and top-3 frequencies of all tested methods under ANC. (d) Rank-1 and top-3 frequencies of all tested methods under collapse threshold.}
    \label{fig7}
\end{figure}

To provide a clearer comparison of the performance of different methods across all real-world hypergraphs, we further take hyper-VDrank as the benchmark and compute the gaps between each baseline method and hyper-VDrank in terms of ANC and collapse threshold, as shown in Figure~\ref{fig7}.
Figure~\ref{fig7}(a) shows the difference between the node-normalized ANC, i.e., \(ANC/N\), of each method and that of hyper-VDrank. Each bar corresponds to one hypergraph, and the normalized ANC differences of different methods are stacked for visualization, with methods having smaller differences placed on the upper level. If the bar segment of a method is located on the right side, it means that its normalized ANC is larger than that of hyper-VDrank, indicating relatively weaker dismantling performance. We can see that, the normalized ANC differences of most baseline methods are located on the right side for most hypergraphs, suggesting that hyper-VDrank generally outperforms the baseline methods. We can also observe that 2-betweenness is a strong baseline and usually ranks second only to hyper-VDrank.
Similarly, Figure~\ref{fig7}(b) shows the difference in the collapse threshold between each method and hyper-VDrank. It is clear that, on most real-world hypergraphs, hyper-VDrank can collapse the hypergraph earlier than most baseline methods. 

It is worth noting that method performance evaluated by ANC and collapse threshold does not change completely synchronously; these two metrics evaluate the effectiveness of hyper-VDrank from different perspectives. For example, a method may reduce the LCC rapidly in the early stage but collapse the hypergraph relatively late, as observed for 2-betweenness on house-committees.
Figure~\ref{fig7}(c) and (d) report the frequency that each method ranks the first or within the top three under the ANC and collapse threshold metrics across different hypergraphs. Hyper-VDrank shows superior performance, achieving the highest frequency in ranking top-1 under both metrics. Moreover, although 2-betweenness is a strong baseline in terms of ANC, its performance in terms of collapse threshold is not clearly superior to EHDF and degree. This indicates that the ability of hyper-VDrank to achieve both higher dismantling efficiency and earlier collapse is nontrivial.

\subsection{Comparison of structural fragmentation during dismantling}
In addition to the size of the largest connected component, i.e., LCC, we further use the number of connected components (CCS) to evaluate the degree of fragmentation caused by different methods. CCS counts how many mutually disconnected parts the residual hypergraph is split into. A larger CCS indicates that the hypergraph is more fragmented. Therefore, LCC and CCS provide two complementary perspectives on the dismantling process: an effective method is expected to reduce LCC rapidly while producing more disconnected components before collapse. 

\begin{figure}
    \centering
    \includegraphics[width=\linewidth]{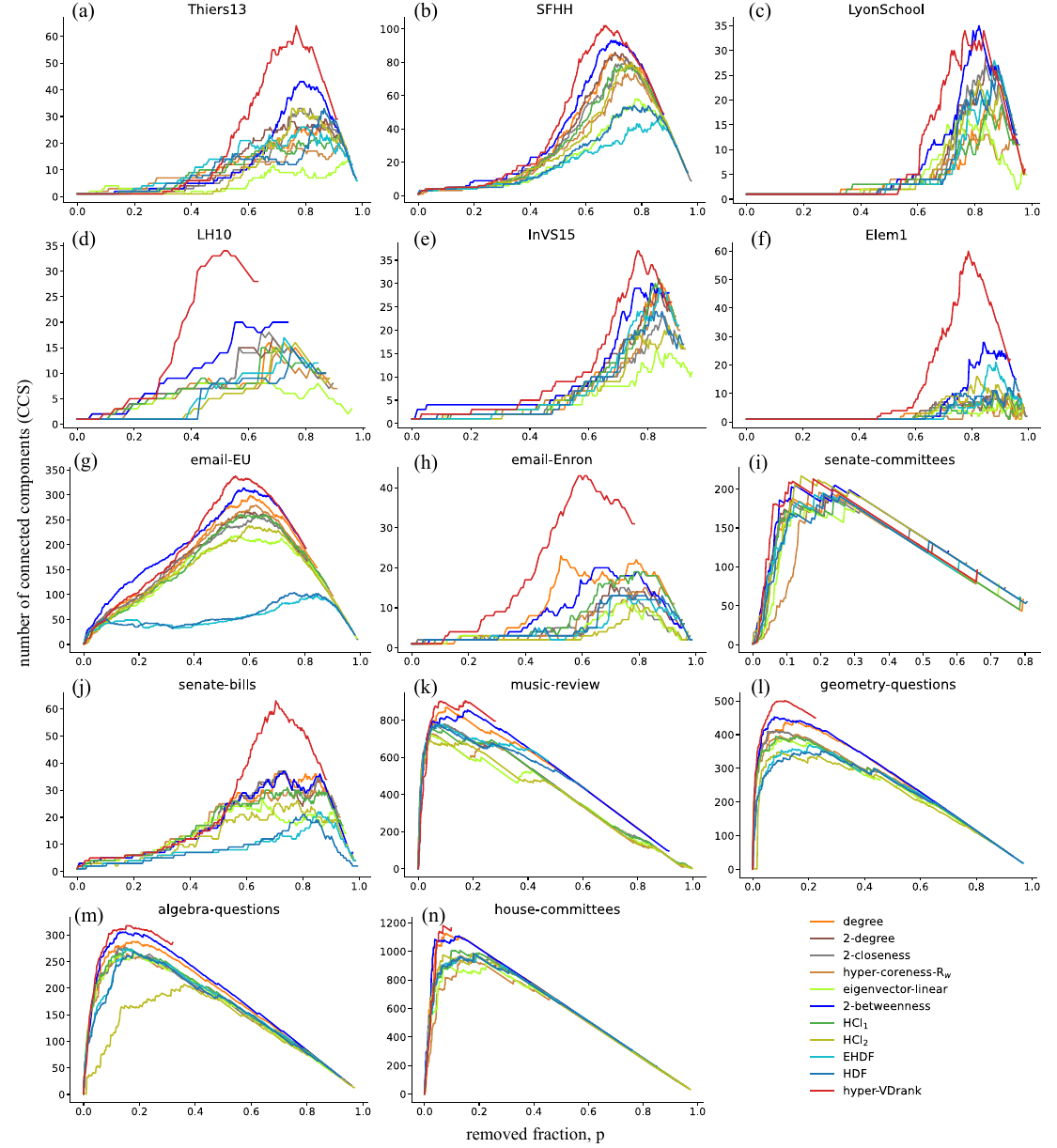}
    \caption{CCS curves on real-world hypergraphs.
    A larger CCS indicates that the residual hypergraph is split into more disconnected parts, reflecting a higher degree of structural fragmentation during dismantling. 
    }
    \label{fig8}
\end{figure}

We plot the CCS curves of different methods during the node removal process on the \(14\) real-world hypergraphs, as shown in Figure~\ref{fig8}. For each method, the CCS curve is truncated when the hypergraph completely collapses, consistent with the LCC curves discussed above. 
It can be seen that the CCS curves of hyper-VDrank are above those of the other methods in most cases, indicating that hyper-VDrank can more effectively split the hypergraph into multiple disconnected components. Together with the LCC results, this suggests that hyper-VDrank not only reduces the largest connected component more rapidly, but also produces a higher degree of structural fragmentation during dismantling. 
We can see that the shown CCS curves have fluctuation and overall first increase and then decrease. This is because, in the early stage, removing structurally critical nodes splits the original large connected component into multiple smaller components, leading to an increase in CCS. 
In the later stage, the isolated nodes generated in earlier steps may be further removed. Since isolated nodes are also counted as connected components, removing such nodes decreases CCS. Thus, the later-stage decrease of CCS does not indicate recovery, but reflects the continued removal of already fragmented residual structures.
Although an ideal dismantling sequence would keep increasing or maintaining a high CCS by continuously attacking non-isolated connected structures, the later-stage decrease of CCS is common for these non-adaptive ranking methods because isolated nodes generated in earlier steps may also appear later in the removal sequence. Nevertheless, hyper-VDrank generally reaches higher CCS values before collapse, verifying its ability to cause a higher degree of structural fragmentation.

\section{Conclusion and discussion}

This paper investigates hypergraph dismantling under strong deletion and proposes hyper-VDrank, a method based on higher-order competition dynamics. The main idea of hyper-VDrank is to identify vulnerable hypergraph structures through node dominance. Specifically, we construct a higher-order information aggregation framework based on environmental feedback between nodes and hyperedges, and design a hyperedge vulnerability weight for the environments in the dynamics. 
We conduct extensive dismantling experiments on 3 synthetic hypegraphs and 14 real-world hypergraphs to evaluate the effectiveness. Results validate that hyper-VDrank achieves higher overall dismantling efficiency, collapses the hypergraph more rapidly, and achieves a higher degree of structural fragmentation. 
The outstanding performance indicates that our hyper-VDrank has a promising potential for risk assessment and intervention in complex systems.

Results suggest that higher-order competition dynamics achieve outstanding performance in hypergraph dismantling. 
The core contribution is how to characterize the information flow and aggregation between nodes and hyperedges, so that higher-order structures and dismantling-rule characteristics can be embedded in node dominance. Compared with simple projection-based pairwise competition, hyper-VDrank preserves higher-order information such as hyperedge size, hyperedge overlap, hyperedge irreplaceability, and establishes the higher-order interaction between nodes and hyperedges, making it more suitable for identifying structurally vulnerable nodes in hypergraph dismantling under strong deletion. More generally, our method provides a new dynamical perspective for understanding and characterizing structural vulnerability in higher-order interaction systems.

Several directions remain for future work. First, this paper adopts a vulnerability weight based on redundancy and the size effect for hyperedges. Future studies may incorporate functional attributes, edge weights, or failure costs in specific systems to design more task-oriented weighting mechanisms~\cite{cost}. Second, this paper focuses on the strong-deletion mechanism, whereas failure processes in real systems may lie between weak deletion and strong deletion. For example, a hyperedge may fail with a certain probability after some of its incident nodes fail, or fail according to a threshold rule~\cite{thresperco}. Extending higher-order competition dynamics to more general higher-order failure rules is therefore a promising direction. Finally, hyper-VDrank is currently mainly used for the structural dismantling of hypergraphs. In the future, it could be further combined with spreading dynamics, cascading failures, and recovery processes to address broader robustness and intervention problems in higher-order systems~\cite{imm,immprr,haounified}.

\section*{Acknowledgements}
This work is supported by National Science and Technology Major Project (2022ZD0116800), Program of National Natural Science Foundation of China (12425114, 12201026, 12501702, 12501718, 62441617), the Fundamental Research Funds for the Central Universities, and Beijing Natural Science Foundation (Z230001), and Beijing Advanced Innovation Center for Future Blockchain and Privacy Computing.

\bibliographystyle{unsrt}
\bibliography{sample}

\clearpage
\appendix

\renewcommand{\thefigure}{A\arabic{figure}}
\setcounter{figure}{0}

\section*{Appendix A \ Search for the optimal competition parameter in real-world hypergraphs}

\begin{figure}[H]
    \centering
    \includegraphics[width=\linewidth]{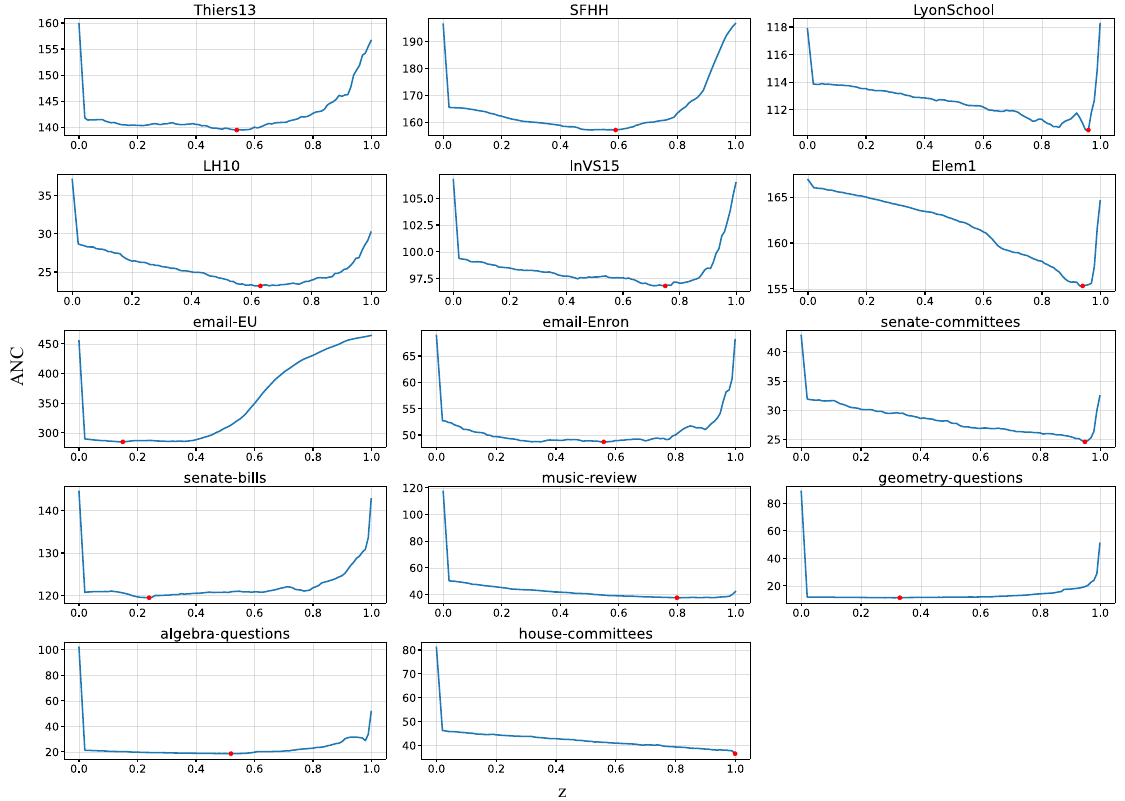}
    \caption{Search process for the optimal competition parameter \(\sigma\) on the \(14\) real-world hypergraphs. The horizontal axis denotes the normalized parameter \(z=-\sigma\lambda_{\min}\), and the vertical axis denotes ANC. The red marker in each subplot indicates the selected optimal parameter that minimizes ANC.}
    \label{figA}
\end{figure}

For hyper-VDrank, the feasible interval of the competition parameter is determined by the minimum eigenvalue of the corresponding higher-order interaction matrix. Specifically, for each real-world hypergraph, we search \(\sigma\) within
\[
\left[0,-\frac{1}{\lambda_{\min}}\right),
\]
where \(\lambda_{\min}\) is the minimum eigenvalue of the matrix \(M\) and \(\sigma=0\) corresponds to a degenerate case with identical node scores and is included as a reference point. To make the search process comparable across different hypergraphs, we use the normalized parameter
\[
z=-\sigma\lambda_{\min}.
\]
Figure~\ref{figA} shows the ANC values obtained under different normalized parameters on the \(14\) real-world hypergraphs. The red marker in each subplot indicates the selected optimal parameter, i.e., the parameter that achieves the minimum ANC.

\clearpage
\renewcommand{\thetable}{B\arabic{table}}
\setcounter{table}{0}
\section*{Appendix B \ Summary of ANC and collapse threshold}

Table~\ref{tab:anc_ct_compare} reports the detailed ANC and collapse threshold \(p_c\) values of all methods on the \(14\) real-world hypergraphs. For both metrics, smaller values indicate better dismantling performance, and the best result on each dataset is highlighted in bold. The last two rows summarize the average relative improvement of hyper-VDrank over each baseline method. Specifically, for a metric \(m\in\{ANC,p_c\}\), the relative improvement of hyper-VDrank over a baseline method \(b\) on dataset \(d\) is computed as
\[
\mathrm{Improvement}_{d,b}^{(m)}
=
\frac{m_{d,b}-m_{d,\mathrm{HVD}}}{m_{d,b}}
\times 100\%,
\]
where \(m_{d,b}\) and \(m_{d,\mathrm{HVD}}\) denote the metric values of baseline method \(b\) and hyper-VDrank on dataset \(d\), respectively. The reported improvement for each baseline is the average over all \(14\) real-world hypergraphs. Averaging over all baseline methods, hyper-VDrank achieves a mean relative improvement of \(23.65\%\) in ANC and \(27.63\%\) in collapse threshold \(p_c\).

\begin{center}
\rotatebox{-90}{%
\begin{minipage}{0.75\textheight}
\centering
\tiny
\setlength{\tabcolsep}{2pt}
\renewcommand{\arraystretch}{0.95}

\captionof{table}{Comparison of ANC and collapse threshold ({$p_c$}) across different methods.}
\label{tab:anc_ct_compare}

\begin{tabular*}{0.75\textheight}{@{\extracolsep{\fill}}llccccccccccc@{}}
\toprule
Dataset & Metric
& HVD
& DC
& 2DC
& 2CC
& HCR
& EL
& 2BC
& HCI$_1$
& HCI$_2$
& EHDF
& HDF \\
\midrule

\multirow{2}{*}{Thiers13}  & ANC & \textbf{139.4434} & 151.8012 & 151.2080 & 148.9878 & 155.0275 & 159.3853 & 145.2966 & 153.1804 & 154.1223 & 152.5963 & 153.1713 \\
         & $p_c$  & \textbf{0.9113} & 0.9450 & 0.9541 & 0.9541 & 0.9541 & 0.9602 & 0.9633 & 0.9541 & 0.9450 & 0.9817 & 0.9786 \\
\midrule

\multirow{2}{*}{SFHH}  & ANC & \textbf{157.0918} & 168.2630 & 167.3896 & 169.8412 & 173.7717 & 180.2035 & 162.0273 & 170.7717 & 172.8238 & 182.1985 & 179.2357 \\
     & $p_c$  & \textbf{0.8759} & 0.9404 & 0.9628 & 0.9777 & 0.9553 & 0.9429 & 0.8784 & 0.9529 & 0.9628 & 0.9256 & 0.9653 \\
\midrule

\multirow{2}{*}{LyonSchool} & ANC & \textbf{110.5165} & 115.9008 & 114.2273 & 114.3802 & 117.0579 & 116.6942 & 112.3802 & 116.7397 & 115.0372 & 115.9421 & 114.9008 \\
           & $p_c$  & 0.9752 & 0.9504 & 0.9587 & 0.9421 & 0.9793 & 0.9793 & 0.9545 & 0.9545 & 0.9380 & \textbf{0.9298} & 0.9463 \\
\midrule

\multirow{2}{*}{LH10} & ANC & \textbf{23.1711} & 31.6447 & 30.7632 & 30.6184 & 32.7500 & 33.5658 & 27.9474 & 31.8026 & 33.6579 & 32.7632 & 33.3421 \\
     & $p_c$  & \textbf{0.6316} & 0.8816 & 0.8947 & 0.8947 & 0.9079 & 0.9605 & 0.7368 & 0.8684 & 0.8816 & 0.8421 & 0.8684 \\
\midrule

\multirow{2}{*}{InVS15} & ANC & \textbf{96.7972} & 101.2581 & 100.6590 & 102.0092 & 101.6037 & 104.0092 & 97.5899 & 100.8802 & 102.7373 & 101.1889 & 101.9908 \\
       & $p_c$  & 0.8802 & 0.9032 & 0.9032 & 0.9263 & 0.9078 & 0.9493 & \textbf{0.8710} & 0.9032 & 0.9263 & 0.9032 & 0.9217 \\
\midrule

\multirow{2}{*}{Elem1} & ANC & \textbf{155.3186} & 168.1032 & 167.1976 & 167.5192 & 169.2920 & 168.0147 & 164.3009 & 168.3156 & 166.7699 & 167.0560 & 166.6903 \\
      & $p_c$  & \textbf{0.9351} & 0.9941 & 0.9941 & 0.9941 & 0.9853 & 0.9823 & 0.9558 & 0.9853 & 0.9823 & 0.9764 & 0.9676 \\
\midrule

\multirow{2}{*}{email-EU} & ANC & \textbf{284.8131} & 310.0276 & 318.3473 & 322.2921 & 320.2993 & 342.1502 & 295.1124 & 326.9346 & 337.9949 & 433.5291 & 428.6925 \\
         & $p_c$  & \textbf{0.8018} & 0.8529 & 0.8815 & 0.9888 & 0.8948 & 0.9867 & 0.8029 & 0.8846 & 0.9734 & 0.9673 & 0.9806 \\
\midrule

\multirow{2}{*}{email-Enron} & ANC & \textbf{48.6993} & 57.9720 & 63.6853 & 64.0420 & 65.8741 & 66.7063 & 58.7273 & 63.5874 & 68.2028 & 65.3147 & 64.1049 \\
            & $p_c$  & \textbf{0.7832} & 0.9161 & 0.9510 & 0.9790 & 0.9231 & 0.9720 & 0.9510 & 0.9580 & 0.9720 & 0.9860 & 0.9860 \\
\midrule

\multirow{2}{*}{senate-committees} & ANC & 24.5745 & 33.5426 & 27.8794 & 26.2872 & 39.1631 & 36.6596 & \textbf{22.6525} & 34.2660 & 24.9752 & 28.9433 & 29.2163 \\
                  & $p_c$  & 0.6596 & 0.8156 & 0.7411 & 0.6348 & 0.7872 & 0.7234 & \textbf{0.5745} & 0.7801 & 0.6702 & 0.7837 & 0.8050 \\
\midrule

\multirow{2}{*}{senate-bills} & ANC & 119.4660 & 122.8605 & 117.3469 & \textbf{117.2891} & 124.1565 & 127.9320 & 119.4320 & 124.1224 & 127.0442 & 139.9830 & 140.3435 \\
             & $p_c$  & \textbf{0.8844} & 0.9150 & 0.9150 & 0.9320 & 0.9320 & 0.9524 & 0.9626 & 0.9422 & 0.9864 & 0.9864 & 0.9932 \\
\midrule

\multirow{2}{*}{music-review} & ANC & \textbf{37.4928} & 44.3668 & 93.3034 & 90.8179 & 120.6495 & 135.0317 & 39.5063 & 100.1721 & 137.6123 & 67.5987 & 74.0444 \\
             & $p_c$  & \textbf{0.2799} & 0.8478 & 0.9937 & 0.9982 & 0.9937 & 0.9982 & 0.9130 & 0.9900 & 0.9964 & 0.4728 & 0.6187 \\
\midrule

\multirow{2}{*}{geometry-questions} & ANC & \textbf{11.3741} & 22.3793 & 41.4310 & 41.0431 & 43.9086 & 52.5379 & 18.8483 & 45.4500 & 66.1655 & 58.6034 & 62.9379 \\
                   & $p_c$  & \textbf{0.2241} & 0.6914 & 0.9069 & 0.8966 & 0.8672 & 0.9052 & 0.8397 & 0.8948 & 0.9241 & 0.9690 & 0.9655 \\
\midrule

\multirow{2}{*}{algebra-questions} & ANC & \textbf{18.6452} & 27.1619 & 34.2000 & 34.3119 & 34.1500 & 35.3238 & 23.4952 & 30.3357 & 78.4786 & 39.3119 & 42.4095 \\
                  & $p_c$  & \textbf{0.3190} & 0.8595 & 0.9119 & 0.9595 & 0.8857 & 0.9452 & 0.9595 & 0.8643 & 0.9714 & 0.9190 & 0.9190 \\
\midrule

\multirow{2}{*}{house-committees} & ANC & 36.5380 & 37.3953 & 76.2442 & 72.4659 & 102.9589 & 96.9558 & \textbf{35.8380} & 65.9891 & 66.6922 & 75.8116 & 76.1496 \\
                 & $p_c$  & \textbf{0.0977} & 0.6295 & 0.9574 & 0.9659 & 0.9558 & 0.9574 & 0.8488 & 0.9457 & 0.9736 & 0.7605 & 0.7597 \\

\midrule
\multirow{2}{*}{\textbf{Improvement}} & \textbf{ANC} & - & \textbf{15.02\%} & \textbf{23.35\%} & \textbf{22.89\%} &\textbf{28.47\%}  & \textbf{29.97\%} & \textbf{7.69\%} & \textbf{25.02\%} & \textbf{28.23\%} & \textbf{27.62\%} & \textbf{28.23\%} \\
                & $\boldsymbol{p_c}$ & - & \textbf{26.40\%} & \textbf{28.05\%} & \textbf{28.24\%} & \textbf{28.46\%} & \textbf{29.74\%} & \textbf{23.42\%} & \textbf{28.14\%} & \textbf{28.61\%} & \textbf{26.67\%} & \textbf{28.57\%}  \\

\bottomrule
\end{tabular*}

\vspace{2mm}
\begin{minipage}{0.75\textheight}
\tiny
\raggedright
HVD: Hyper-VDrank; DC: degree centrality; 2DC: 2-degree centrality; 
2CC: 2-closeness centrality; HCR: hyper-coreness-R$_w$; 
EL: eigenvector-linear; 2BC: 2-betweenness centrality.
\end{minipage}

\end{minipage}%
}
\end{center}


\end{document}